\documentclass[aps,pre,
twocolumn,
groupedaddress,showpacs]{revtex4}
\usepackage{graphicx}
\usepackage{epstopdf}
\begin{document}


\title{Farey sequence in the appearance of subharmonic Shapiro steps
}
\author{Jovan Odavi\' c$^{1}$, Petar Mali$^{2}$, and Jasmina Teki\' c$^{3}$}
\affiliation{$^1$ Institut f\"ur Theorie der Statistishen Physik - RWTH Aachen University,
Peter-Gr\"unberg Institut and Institute for Advanced Simulation, Forschungszentrum J\"ulich, Germany,}
\affiliation{$^2$ Department of Physics, Faculty of Science, University of Novi Sad,
Trg Dositeja Obradovi\' ca 4, 21000 Novi Sad, Serbia}
\affiliation{$^3$"Vin\v ca" Institute of Nuclear Sciences,
Laboratory for Theoretical and Condensed Matter Physics - 020,
University of Belgrade, PO Box 522, 11001 Belgrade, Serbia}
\date{\today}


\begin{abstract}
The largest Lyapunov exponent has been examined in the dynamical-mode locking
phenomena of the ac+dc driven dissipative Frenkel-Kontorova model with deformable substrate potential.
Due to deformation, large fractional and higher order subharmonic steps appear in the response function of the system.
Computation of the largest Lyapunov exponent as a way to verify their
presence led to the observation of the Farey sequence.
In the standard regime, the appearance of halfinteger and other subharmonic steps between the large harmonic steps,
and their relative sizes follow the Farey construction.
In the nonstandard regime, however, the halfinteger steps are larger than harmonic ones, and
Farey construction is only present in the appearance of higher order subharmonic steps.
The examination of Lyapunov exponents has also shown that regardless of the substrate potential or deformation,
there was no chaos in the system.

\end{abstract}
\maketitle

\section{Introduction}
\label{intro}

In the examination of Shapiro steps, finding the best method or tool
to verify their presence has been the matter of many studies in
various physical systems.
Numerous theoretical and experimental results on Shapiro steps obtained in
dissipative systems such as charge- or spin-density waves conductors \cite{Grun, Thorn, McC, SSC},
the systems of Josephson junction arrays \cite{Benz, Sohn, Dub} and
superconducting nanowires \cite{Dins, Bae} have been the main impulse and motivation
for our studies of the ac+dc driven overdamped (dissipative) Frenkel-Kontorova (FK) model \cite{OBBook}.
It is well known that when these systems are subjected under an
external ac driver, their dynamics is characterized by the appearance of  Shapiro steps.
These steps are due to interference phenomena or dynamical mode-locking (synchronization) of the
internal frequency with the applied external one. If the locking
appears at the integer values of the external frequency, the steps
are called harmonic while for the locking at rational (noninteger)
values of frequency they are called subharmonic.

In realistic systems due to presence of noise, impurities and other environmental effects,
detection of Shapiro steps, particularly the subharmonic ones, is usually very difficult.
On the other hand, in theoretical works, their observation could also be a problem since their size is often so small
that they are invisible on the regular plot of the response function.
One of the most sensitive ways to verify the existence of Shapiro steps is the calculation of {\it the largest Lyapunov exponent} \cite{Hilborn}.
Always when the system is dynamically mode-locked,
the largest Lyapunov exponent has negative values \cite{FlorAP, Falo}.
Therefore, an examination of the largest Lyapunov exponent for some interval of driven force will precisely reveal
the presence of any harmonic or subharmonic mode-locking.

Calculation of the largest Lyapunov exponent has been already used as a way to examine the existence of subharmonic Shapiro steps in
the standard FK model \cite{FlorAP, Falo}.
The standard Frenkel-Kontorova (FK) model represents a chain of harmonically interacting
particles subjected to a sinusoidal substrate potential \cite{OBBook}.
It describes different commensurate or incommensurate structures
that under an external driving force, show very rich dynamical behavior.
In the presence of an external ac+dc driving force, the dynamics
is characterized by the appearance of the staircase macroscopic response
or the Shapiro steps in the response function $\bar v(\bar F)$ of the system \cite{FlorAP, Falo, Flor}.
Though the standard FK model has been very successful in the studies of some effects
related to Shapiro steps, its applications is still very restricted.
Namely, in the standard FK model, the subharmonic steps either do not exists in
the case of commensurate structures with integer values of winding
number \cite{Renne, Wald} or their size is so small that analysis of
their properties is very difficult \cite{FlorAP, Falo, Flor}.
The absence of subharmonic steps for the commensurate structures with integer value of winding number, and
their presence in the case of rational (noninteger) winding number was confirmed by the calculation of the largest Lyapunov exponent \cite{Falo}.
However, contrary to the standard case, the large subharmonic steps can appear in any commensurate structure of the nonstandard FK model
such as the one with the asymmetric deformable substrate potential (ASDP) \cite{AC2}.
This potential belongs to the family of nonlinear periodic deformable potentials, introduced by Remoissent and Peyrard \cite{Peyr}
as the way to model many specific physical situations without employing perturbation methods.

In this paper, by using the largest Lyapunov exponent computation technique, we will examine the appearance of both harmonic and subharmonic steps
in the FK model with asymmetric deformable substrate potential (ASDP).
In the analysis of the largest Lyapunov exponent, we have observed one interesting property:
the Shapiro steps and their relative sizes appear according to the {\it Farey construction} only in the standard regime
when large harmonic steps are dominant in the response function.
The paper is organizes as follows.
The model and methods are introduced in Sec. II, and the results
are discussed in Sec. III.
Finally, Sec. IV concludes the paper.

\section{Model and methods}
\label{model}

\begin{figure}[ht] 
\includegraphics[width=6.0cm]{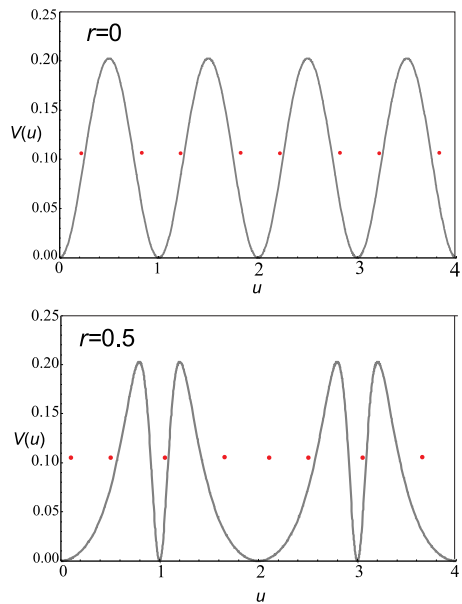}
\centering
\caption{\label{Fig1}
(Color online) Particles moving in asymmetric deformable potential for $\omega =\frac 12, K=4$, and two
different values of the shape parameters $r=0$ and $r=0.5$. Particles are represented by red dots.}
\end{figure}

We consider the dissipative (overdamped) dynamics of a series of coupled harmonic
oscillators $u_l$ subjected in the ASDP:
\begin{equation}
\label{Vr}
V(u)=\frac {K}{(2\pi )^2}\frac {(1-r^2)^2\left[
1 - \cos (2\pi u)\right] } {\left[ 1 + r^2 + 2r\cos (\pi
u)\right] ^2},
\end{equation}
where $K$ is the pinning strength, and $r$ is the shape or deformation parameter ($-1< r <1)$ which can be varied continuously.
By changing  the shape parameter $r$ the potential can be tuned in a very fine way, from the simple sinusoidal one for $r=0$ to deformable for $0<r<1$.
In Fig \ref{Fig1}, the commensurate structure $\omega=1/2$ in ASDP is presented  for two different values of the shape parameter $r=0$ and $r=0.5$
(for more details see \cite{AC2, ACr}).
The average interparticle distance $\omega=\langle u_{l+1}-u_l \rangle$, or the so called winding number is
one of the main parameters that describes the FK model.
The system exhibits commensurate phase for rational values of winding number $\omega$, and incommensurate phase
for irrational ones.

In the present paper, the system of coupled harmonic oscillators in ASDP is driven by the dc+ac forces
$F(t) =\bar F + F _{ac}\cos (2\pi \nu _{0}t)$.
The equations of motion are
\begin{equation}
\label{u}
\dot u_l
=u_{l+1}+u_{l-1}-2u_{l}-\frac{\partial V}{\partial u_l}+F(t),
\end{equation}
where $l=-\frac N2,...,\frac N2$, $u_l$ is the position of $l$th particle,
$\bar{F}$ is dc force, where $F_{ac}$ and $2\pi\nu_0$ are the amplitude and the circular frequency of the ac force, respectively.
Since the substrate potential is homogeneous (it does not depend of the particles index $l$) relabeling of the position of particles will not change
the properties of the configuration \cite{OBBook}.
When the system is driven by an external ac+dc force, two different frequency scales appear in the system:
the frequency of the external periodic force $\nu_0$, and the characteristic frequency of the motion of particles over the ASDP driven by the
average force $\bar{F}$.
The competition between those frequency scales can result in the appearance of resonance (dynamical mode locking or Shapiro steps).

Solution of the system (\ref{u}) is called resonant if average velocity $\bar{\upsilon}$ satisfies the relation:
\begin{equation}
\label{vel} \bar{v}= \frac {i\omega + j a}{m} \nu_0,
\end{equation}
where triplet ($i,j,m$) are integers numbers.
Resonant velocity is called harmonic if $m=1$ and subharmonic if $m\neq 1$. (In case of $\omega=\frac{1}{q}$ we can use $\bar{v}=\frac{i }{m}\omega\nu_0$ for marking harmonic and subharmonic steps). 
Parameter $a$ is the period of the potential $V(u)$ and
in the case of no deformation $a=1$, and with deformation $a=2$ as can be seen in Fig \ref{Fig1}.
For a rational value of $\omega=p/q$ ($p$ and $q$ coprime integers) the triplet is not unique
(this triplet is unique only for incommensurate structures \cite{OBBook}).
In this paper we will consider only the commensurate structure $\omega = 1/2$.

The equations of motion (\ref{u}) have been integrated by using a fourth order Runge-Kutta method with periodic boundary conditions.
The time step used in simulations was $0.01\tau$, where $\tau=\frac{1}{\nu_0}$.
The force is varied adiabatically with the step $10^{-5}$.

We shall be focused on calculating the largest Lyapunov
exponent $\lambda$ \cite{Falo}.
It is well known that the Lyapunov exponent gives a quantitative measure on the presence
of chaos in dynamical systems \cite{Hilborn}, however,
it also proves to be extremely sensitive to the existence of both harmonic and subharmonic Shapiro steps.
When the system is dynamically mode locked, i.e. on the step, the trajectories of particles are periodic in time
which is reflected by the negative value of the largest Lyapunov exponent.
Outside the steps, where there is no onset of dynamical mode locking,
the trajectories are quasi-periodic which is confirmed by the zero of
the Lyapunov exponent (\cite{Hilborn, Falo}).
We choose an appropriate perturbed point $u_l'$ in our computations according to:
\begin{equation}
\label{Lap} u_{l}' (t_{ss}) = u_{l} (t_{ss}) \pm \sqrt{\frac{d_0^2}{N}}
\end{equation}
where $t_{ss}$ is time when the steady-state has been achieved in our system, and
$d_0$ is the parameter that expresses the change in the initial positions of particles of the model.
In order to make sure that projecting is always done onto the subspace dominated by the largest Lyapunov exponent,
the sign in front of the square term in Eq.\ref{Lap} is randomly selected where the plus and minus sign appear
with the same probability.
We sample and readjust following Sprott \cite{Sprott} every 25 or so time steps.
In our calculations, we used $t_{ss}=300\tau$ and $d_0 = 10^{-7}$.
For convenience, in further text, the largest Lyapunov exponent will be denoted just as the Lyapunov exponent.

\section{Results}
\label{results}

\begin{figure}[ht] 
\includegraphics[width=5.8cm]{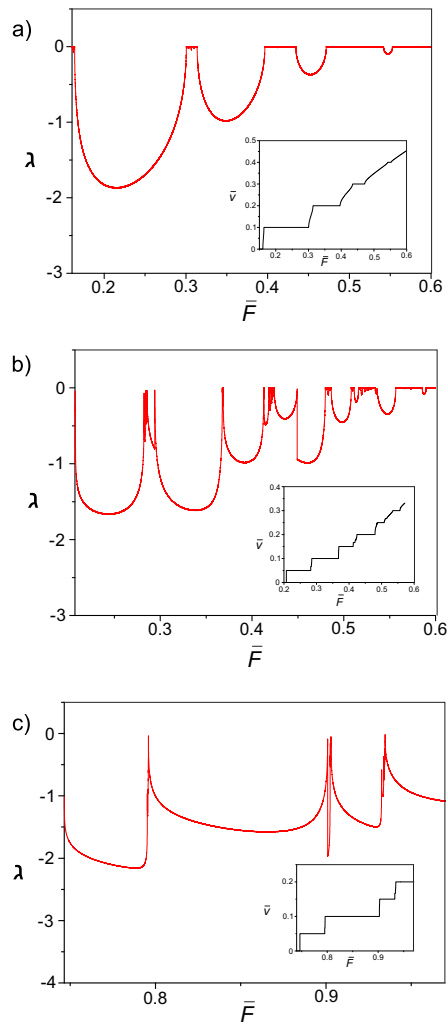}
\centering
\caption{\label{Fig2}
(Color online) The Lyapunov exponent as a function of the average force for commensurate structure
$\omega =1/2, K=4, F_{ac}=0.2, \nu _0=0.2 $ and
three different values of shape parameter (a) $r=0$, (b) $r=0.28$ and (c) $r=0.6$.
The insets show the corresponding response functions $\bar v(\bar F)$.}
\end{figure}

In the present paper, the Lyapunov exponent is examined for different deformations of the substrate potential.
In Fig. \ref{Fig2}, the Lyapunov exponent as a function of the driving force for three different values of deformation
parameter is presented.
The insets show the corresponding response functions $\bar v(\bar F)$ (the average velocity as a function of average driving force).
As one can see, the computed Lyapunov exponents are always $\lambda \leq 0$,
which implies that with the change of deformation $r$ we have not introduced chaos in our system
(presence of chaos would result in the positive values of Lyapunov exponent).
Domain of $\bar{F}$ in Fig. \ref{Fig2}, for which we calculated the exponent, differs with $r$
due to the fact that for different values of $r$ the same steps appear in different regions of $\bar{F}$ (see \cite{AC2, ACr}).
In the standard case in Fig. \ref{Fig2} (a), we can see the large minima which correspond to harmonic steps and which size changes monotonically.
As deformation increases in Fig. \ref{Fig2} (b) and (c), the minima which corresponds to the large halfinteger and higher order subharmonic steps appear
where the changing of their size is not monotonic any more.

\begin{figure}[ht] 
\includegraphics[width=6.0cm]{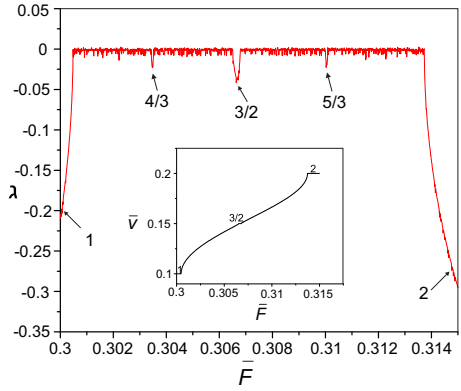}
\centering
\caption{\label{Fig3}
(Color online) Lyapunov exponent as a function of average driving force for $r=0$ (the rest of parameters are as in Fig. \ref{Fig2}).
The inset shows the response function $\bar{v}(\bar{F})$ drawn for the same interval of force.
This result is obtained in \cite{Falo}.}
\end{figure}

Using the Eq.(\ref{vel}), the Shapiro steps could be now identified.
It is well known that in the standard FK model ($r=0$) with integer value of winding number, there would be no subharmonic mode locking \cite{OBBook},
and consequently, no steps between harmonic ones on the plot of response function $\bar{v}(\bar{F})$.
On the other hand, when $\omega =1/2$, only halfinteger step $3/2$ which appears between the first and the second harmonic
could be visible \cite{Falo}.
However, computation of the Lyapunov exponent between first and the second harmonic steps reveal other subharmonic steps as can be seen in
Fig. \ref{Fig3}.
The areas under the minima correspond to the size of the steps, i.e. for larger step, the area under the minimum will be larger.
If we examine the subharmonic steps in Fig. \ref{Fig3}, we could see that the first largest fractional step between
the step $1$ and the step $2$ is the step $3/2$.
Then, the largest step between the steps $1$ and $3/2$ would be the step $4/3$ while the largest one between the steps $3/2$ and $2$ is $5/3$.
Therefore, according to the appearance of fractional steps between the first $1/1$ and the second $2/1$ harmonics we may write the following sequence:

\begin{equation}
\frac{1}{1}, \frac{4}{3},\frac{3}{2},\frac{5}{3},\frac{2}{1}.
\end{equation}

This sequence of numbers represents the {\it Farey sequence} well known in number theory \cite{Thom, Hardy}.

\begin{figure}[ht] 
\includegraphics[width=6.0cm]{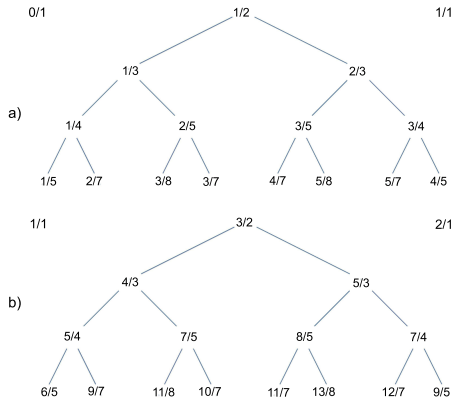}
\centering
\caption{\label{Fig4}
(Color online) Section of the Farey construction (represented as a rooted binary tree graph) a) from $0$ to $1$, b) from $1$ to $2$ according to Eq. \ref{far}. }
\end{figure}

The Farey sequence $\mathcal{F}_{\mathcal{N}}$ of order $\mathcal{N}$  is
an ascending sequence of irreducible fractions between $0$ and $1$, whose denominators are less or equal then $\mathcal{N}$ \cite{Thom, Hardy}.
The first few would be:
\begin{equation}
\label{FN}
\begin{array}{l l}
\mathcal{F}_{1}& =\Big\{\frac{0}{1}, \frac{1}{1}\Big\}\\
\mathcal{F}_{2}& =\Big\{\frac{0}{1},\frac{1}{2}, \frac{1}{1}\Big\}\\
\mathcal{F}_{3}& =\Big\{\frac{0}{1}, \frac{1}{3},\frac{1}{2},\frac{2}{3},\frac{1}{1}\Big\}\\
\mathcal{F}_{4}& =\Big\{\frac{0}{1}, \frac{1}{4}, \frac{1}{3},\frac{1}{2},\frac{2}{3},\frac{3}{4},\frac{1}{1}\Big\}\\
\mathcal{F}_{5}& =\Big\{\frac{0}{1},\frac{1}{5}, \frac{1}{4}, \frac{1}{3},\frac{2}{5},\frac{1}{2},\frac{3}{5},\frac{2}{3},\frac{3}{4},\frac{4}{5},\frac{1}{1}\Big\}
\end{array}
\end{equation}
Therefore, if we have two rational fractions in Farey sequence $\frac{p}{q}$ ($p$, $q$ are coprime integers) and $\frac{p'}{q'}$ ($p'$,$q'$ are coprime integers),
the rational fraction which lies between them and has the smallest denominator is
\begin{equation}
\label{far}
\frac{p''}{q''}=\frac{p}{q}\oplus \frac{p'}{q'}=\frac{p+p'}{q+q'}
\end{equation}
where $p''$, $q''$ are coprime integers. This statement is trivially
extended to the case of interval between $1$ and $2$, and further on
(Theorems $28$ and $29$ in \cite{Hardy}). The largest step between
$\frac{p}{q}$ and $\frac{p'}{q'}$, if exists, will be step
$\frac{p}{q} \oplus \frac{p'}{q'}$.
The Farey sequence could be easily understood from the diagram in Fig. \ref{Fig4}

For example, in the case of the FK model with the integer value of winding number, there is no subharmonic mode locking,
which implies there is only the Farey sequence of order one.
However, if the winding number is rational noninteger such as the case $\omega =1/2$ in Fig. \ref{Fig3},
one can see that the largest step between harmonic steps $1$ and $2$ is halfinteger step $3/2$.
From the set theory \cite{Hardy} we know that between any two rational fractions lie countable many, $\aleph_0$ rational fractions and
therefore, countable many possible Shapiro steps between any two harmonic steps in our model.

\begin{figure}[ht] 
\includegraphics[width=6.0cm]{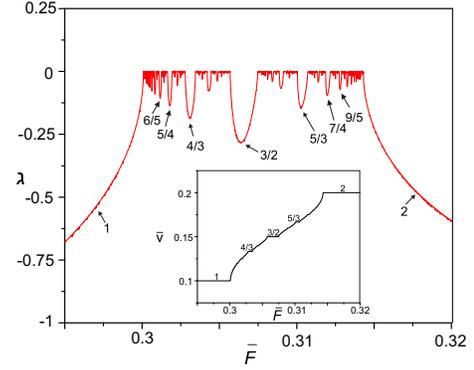}
\centering \caption{\label{Fig5}
(Color online) Lyapunov exponent as a function of average driving force for $r=0.01$ (the rest of parameters are as in Fig. \ref{Fig2}).
The inset shows the response function $\bar{v}(\bar{F})$ drawn for the same interval of
force.}
\end{figure}

If the potential gets deformed, the large halfinteger step and higher order subharmonic steps appear \cite{AC2, ACr}.
Contrary to the case $r=0$ in Fig. \ref{Fig3}, now for $r=0.01$ in Fig. \ref{Fig5}, the large $4/3$ and $5/3$ steps are clearly visible.
The higher order subharmonic steps, such as $4/3$ and $5/3$ (to the left and to the right),
are appearing in a symmetric manner with respect to the step $3/2$.

\begin{figure}[ht] 
\includegraphics[width=6.0cm]{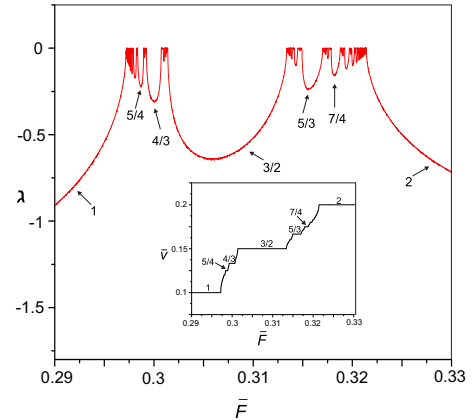}
\centering
\caption{\label{Fig6}
(Color online) Lyapunov exponent as a function of average driving force for $r=0.05$ (the rest of parameters are as in Fig. \ref{Fig2}).
The inset shows the response function $\bar{v}(\bar{F})$ drawn for the same interval of force.}
\end{figure}

With the further increase of deformation $r$, the step widths increase faster on the right side from halfinteger step $3/2$
than on the left one as can be seen in Fig. \ref{Fig6}.

\begin{figure}[ht] 
\includegraphics[width=6.0cm]{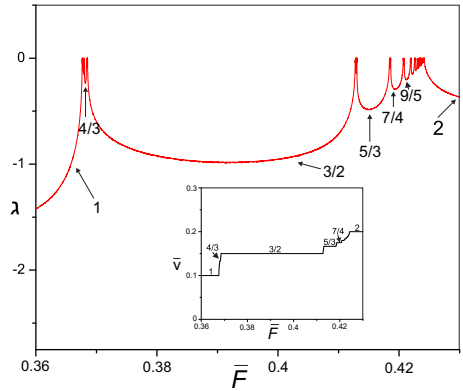}
\centering
\caption{\label{Fig7}
(Color online) Lyapunov exponent as a function of average driving force for $r=0.28$ (the rest of parameters are as in Fig. \ref{Fig2}).
The inset shows the response function $\bar{v}(\bar{F})$ drawn for the same interval of force.}
\end{figure}

It was shown in our previous work \cite{ACr}, that size of halfinteger and subharmonic Shapiro steps increase with deformation and
after reaching their maxima for some value of $r$, decrease to zero.
If we calculate the Lyapunov exponent for $r=0.28$, which is the value of $r$ for which halfinteger step reaches its maximum,
we obtain results presented in Fig. \ref{Fig7}.
At this value of deformation some higher subharmonic steps already start to disappear.

\begin{figure}[ht] 
\includegraphics[width=6.0cm]{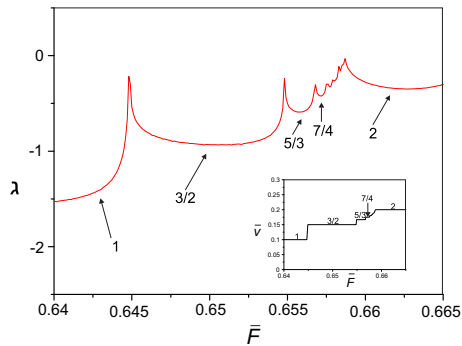}
\centering
\caption{\label{Fig8}
(Color online) Lyapunov exponent as a function of average driving force for $r=0.5$ (the rest of parameters are as in Fig. \ref{Fig2}).
The inset shows the response function $\bar{v}(\bar{F})$ drawn for the same interval of force.}
\end{figure}

At large deformation of the potential,
the size of halfinteger steps decreases, and the most of higher order subharmonic steps have completely vanished \cite{ACr}.
This is confirmed by the results in Fig. \ref{Fig8}, where the Lyapunov exponent for $r=0.5$ has been calculated.
Disappearance of steps is also clearly visible in Fig. \ref{Fig2}.

Deformation of the potential obviously strongly affects the steps as we can see in Fig. \ref{Fig6}-\ref{Fig8}.
It appears that  with the increase of the deformation $r$, the right side of the Farey construction is heavily favored over the left one.
In particular, we observe that at each level of our binary tree graph (Farey construction in Fig. \ref{Fig4}) the child node (or step)
that takes preference is the one on the right.
This means that with the increase of the deformation the steps that are present and become increasingly dominant are $3/2,5/3,7/4,9/5$.

We have analyzed also the systems with other types of deformable potentials \cite{ACP}, such as variable, double barrier and double well potential, and
we have been always able to observe the appearance of steps in accordance with the Farey construction \cite{Hilborn}.
Therefore, for two steps $\frac{p}{q}$ and $\frac{p'}{q'}$, the next largest step between them will be $\frac{p+p'}{q+q'}$ where
denominator determines the size of steps in terms that the size of steps decreases as the denominator increases.
It is important to note that Farey construction tell us the order and the relative sizes of steps but
it does not tell us the actual step width or why they appear \cite{Hilborn}.

It is well known that the size of harmonic and halfinteger steps are correlated, whereby the larger the size of harmonic
the smaller the one of halfinteger step and vice versa \cite{JLee, Earl, ACr}.
In some cases, depending on the system parameters, the size of halfinteger steps could be even larger than the size of harmonic ones \cite{ACr}.
The size of halfinteger and other subharmonic steps strongly affects the behavior of the system, and according to that
in previous works \cite{ACr, Earl}, the three different types of system behavior have been classified: the standard behavior for small halfinteger steps,
the behavior for intermediate halfinteger steps and the behavior in the presence of large halfinteger steps.

\begin{figure}[ht] 
\includegraphics[width=6.0cm]{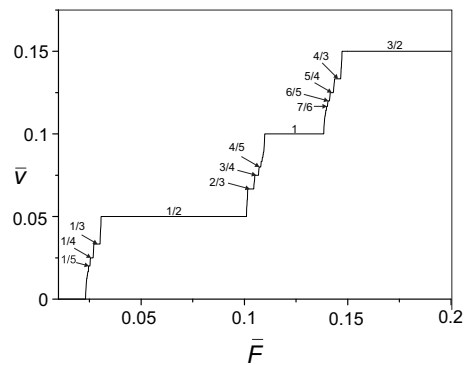}
\centering
\caption{\label{Fig9}
The average velocity as a function of average driving force for $F_{ac}=0.55$, and $r=0.2$ (the rest of parameters are as in Fig. \ref{Fig2}).
The numbers mark halfinteger and subharmonic steps.}
\end{figure}

If we have two harmonic steps, according to Farey sequence the next largest step which appears between them is the halfinteger step,
but this is not the case for nonstandard behavior \cite{ACr, Earl}, since halfinteger steps are larger than harmonic ones.
In such case, could we still have the presence of Farey sequence?
In Fig. \ref{Fig9}, the response function in the case of large halfinteger steps is presented.
It is obvious that in the nonstandard case, the relative sizes of harmonic and halfinteger steps do not follow the Farey construction,
and going from harmonic to halfinteger steps, the size of step does not decreases as denominator increases, on the contrary,
the halfinteger steps $1/2$ and $3/2$ are larger than harmonic ones $1/1$ and $2/1$.
However, the higher order subharmonic steps between halfinteger and harmonic steps
still appear according to Farey construction and their sizes decrease as the denominator increases.

Calculation of the Lyapunov exponent gives a possibility not only to detect all resonances in the response function, but
also to detect the presence of chaos.
In all our simulations performed on the ac+dc driven overdamped FK model we did not observed any chaos.
Presence of deformable substrate potentials and different level of deformations did not introduced chaotic behavior into the system.
Contrary to our case, chaos has been observed in the spatiotemporal dynamics of moving kinks in the damped dc driven FK model
where the resonances appear due to competitions between the moving kinks and their radiated phase modes \cite{ZHuHu}.
Also, structured chaos has been observed in a Josephson junction systems where chaotic regions appear between the subharmonic Shapiro steps
at certain values of system parameters \cite{Shukr}.

\section{Conclusion}
\label{concl}

In this paper, we have presented  detailed analysis of the Shapiro steps in the ac+dc driven dissipative FK model
by using the Lyapunov computation technique.
The obtained results show the presence of Farey sequence in the appearance of subharmonic steps.
The steps and their relative sizes follow exactly the Farey construction only in the standard regime when
harmonic steps are the largest one.
However, in the nonstandard regime, the halfinteger steps are larger than harmonic ones, and
Farey sequence appears only in the order and relative sizes of higher order subharmonic Shapiro steps.
Lyapunov exponent analysis is certainly one  of the best ways to get an accurate answer about the presence of chaos in the system.
Computations of Lyapunov exponent have been performed for different system parameters, and
regardless of the deformation, no chaos has been observed in the behavior of the system.
Absence of chaos in the presence of deformable potentials certainly requires further investigation.
This problem and the possibility of chaotic behavior in other situations such as presence of noise
will be subject of future examinations.

Presented results could be important for the studies of Shapiro steps in all systems that are closely related
to the dissipative dynamics of the FK model.
In experimental and theoretical works performed in charge density wave systems and the systems of Josephson junction arrays,
measuring  of differential resistance is usually used to detect subharmonic steps.
If we look for example at the results obtained in sliding charge-density wave systems \cite{Thorn},
the systems of Josephson junction arrays \cite{Sohn, Dub} or superconduction nanowires \cite{Dins},
we can observe the presence of Farey construction in the appearance of Shapiro steps.
Our analysis shows that Farey construction can not be always generally applied when it comes to relative sizes of the observed steps.
Since the appearance and origin of the subharmonic Shapiro steps are still a matter of debates, we hope that these
results could bring an insight into understanding of these physical phenomena.


\begin{acknowledgments}
We wish to express our gratitude to Prof. P. J. Mart\' inez for helpful discussions.
This work was supported by the Serbian Ministry of
Education and Science under Contracts No. OI-171009 and No. III-45010 .

\end{acknowledgments}



\end{document}